\documentclass[conference,compsoc]{IEEEtran}
\IEEEoverridecommandlockouts
\usepackage{color}
\usepackage{soul}
\usepackage{url}  % Formatting web addresses  
\usepackage{ifthen}  % Conditional 
\usepackage{multicol}   %Columns
\usepackage[utf8]{inputenc} %unicode support
\usepackage{graphics}
\usepackage{graphicx}
\usepackage{amsmath,amssymb,amsfonts}
\usepackage{algorithmic}
\usepackage{textcomp}
\usepackage{xcolor}
\usepackage{epsfig,epstopdf}
\usepackage[nocompress]{cite}

\def\BibTeX{{\rm B\kern-.05em{\sc i\kern-.025em b}\kern-.08em
    T\kern-.1667em\lower.7ex\hbox{E}\kern-.125emX}}

\usepackage[pscoord]{eso-pic}

\begin{document}

\title{GSSMD: A new standardized effect size measure to improve robustness and interpretability in biological applications}

\author{\IEEEauthorblockN{Seongyong Park and Shujaat Khan}
\IEEEauthorblockA{\textit{Department of Bio and Brain Engineering} \\
\textit{Korea Advanced Institute of Science and Technology}\\
Daejeon, South Korea \\
\{sypark0215, shujaat\}@kaist.ac.kr}
\and
\IEEEauthorblockN{Muhammad Moinuddin and Ubaid M. Al-Saggaf}
\IEEEauthorblockA{\textit{Center of Excellence in Intelligent Engineering Systems} \\
\textit{King Abdulaziz University}\\
Jeddah, Saudi Arabia. \\
\{mmsansari, usaggaf\}@kau.edu.sa}
}
\maketitle

\begin{abstract}
In many biological applications, the primary objective of study is to quantify the magnitude of treatment effect between two groups. Cohens’d or strictly standardized mean difference (SSMD) can be used to measure effect size however, it is sensitive to violation of assumption of normality. Here, we propose an alternative metric of standardized effect size measure to improve robustness and interpretability, based on the overlap between two sample distributions. The proposed method is a non-parametric generalized variant of SSMD (Strictly Standardized Mean Difference). We characterized proposed measure in various simulation settings to illustrate its behavior. We also investigated finite sample properties on the estimation of effect size and draw some guidelines. As a case study, we applied our measure for hit selection problem in an RNAi experiment and showed superiority of proposed method. 
\end{abstract}


\section{Introduction}
In typical biological experiments, we often divide independent groups into controls and cases to quantify the treatment effect. The effect size is a statistical metric that provides an estimate of the difference between the two groups, and it is an indicator that is calculated as a standardized mean difference (SMD) only if the population parameters are known. 

Cohen’s $d$ can be used to estimate the population effect size based on sample distributions and it is defined as sample mean difference divided by pooled estimate of sample standard deviation. Variants of Cohen's $d$ such as Glass's $\delta$ and Hedge's $g$ can be used for the same purpose and they have the same form but use slightly different way to estimate the sample variance. The strictly standardized mean difference (SSMD)\cite{zhang2007pair} is another variant of Cohen's $d$ and it is popular in bioassay quality control and hit selection problems. Different bioassay quality control measures for effect size estimation are available e.g., Z'-factor and SSMD, and robust SSMD etc. The primary difference in robust variants is that they utilize median and MAD instead of mean and variance.

The drawbacks of aforementioned effect size measures are sensitivity toward underlying distribution and interpretability. Since these measures are only utilize statistical parameters such as sample mean and variance, they are susceptible to noise in the measurement or distribution transformation. In addition, these measures can be difficult to interpret since the range of values are $(-\infty, \infty)$ and due to the distributional differences and measurement noise the meaning of the value changes with application domain.

To address aforementioned issues, we propose a new effect size measure called generalized strictly standardized mean difference (GSSMD). Since the GSSMD is defined by the non-overlap proportion between two distributions, it avoids the complexity of other effect size measure calculations in non-standard or transformed distributions. By definition, it is intuitive enough and easy to interpret. In this paper, we used simulations and real biological data sets to support our claims. Mathematical definitions and experimental results are described in the following sections.

\section{Methods}

\subsection{Definition of overlap statistics and GSSMD}
In general, distribution overlap between two continuous probability density functions (PDFs) is defined as 

\begin{equation}
OVL = \int_{R^n}\min[f_A(x), f_B(x)] dx
\end{equation}

where $f_A (x)$ and $f_B (x)$ represent two sample distributions. If samples are drawn from two normal distributions $f_{A}(x)$ and $f_{B}(x)$, then $OVL$ is equivalent to 

\begin{equation}
OVL = \int_{R^n}\min[N(x|\bar{x}_{A},s^2_{A}) , N(x|\bar{x}_{B},s^2_{B})] dx,
\end{equation}

This is equivalent to 

\begin{equation}
OVL = 2\Phi\bigg(-\frac{|\bar{x}_{pos} - \bar{x}_{neg}|}{2s_{pooled}}\bigg),
\end{equation}
where, $\Phi\left(\cdot\right)$ is the cumulative distribution function. To use overlap statistic as differential measure, herein we defined GSSMD as

\begin{equation}
GSSMD=sign\left({\bar{x}}_{pos}-{\bar{x}}_{neg}\right)\times\left(1-OVL\right)
\end{equation} where $sign(\cdot)$ is sign operator and the GSSMD is in the range $-1\le GSSMD \le1$.

\subsection{GSSMD and detection probability}
Let us assume two probability density functions $P_{pos}\left(x\right)$ and $P_{neg}\left(x\right)$ representing positive and negative population distributions of an assay and a random variable $x$ representing assay values as shown in supplementary Figure S1. Here we assumed that $\mu_{pos}>\mu_{neg}$. The type I (False Positive Rate, FPR) and type II (False Negative Rate, FNR) errors or mis-classification risks of a classifier $h\left(x\right)$ with a given threshold $\epsilon$ can be formulated by

$R_0\left(h\middle|\epsilon\right)=P\left(h\left(x|\epsilon\right)\neq Y\middle| Y=0\right)$ and

$R_1\left(h\middle|\epsilon\right)=P\left(h\left(x|\epsilon\right)\neq Y\middle| Y=1\right)$.

The formulae can be obtained by integrating two distributions

$R_0\left(h|\epsilon\right)=\int_{\epsilon}^{\infty}{p_{neg}\left(x\right)}dx$ and

$R_1\left(h\middle|\epsilon\right)=\int_{-\infty}^{\epsilon}{p_{pos}\left(x\right)}dx$.

Hence, a probability of error or total risk of mis-classification can be calculated as 

\begin{equation}
\begin{split}
R\left(h\middle|\epsilon\right)&=P\left(h\left(x|\epsilon\right)\neq Y\right) \\
&=P\left(Y=0\right)R_0\left(h|\epsilon\right)+P\left(Y=1\right)R_1\left(h|\epsilon\right)
\end{split}
\end{equation}

If the prior probability $P\left(Y=0\right)=P\left(Y=1\right)=0.5$, then the risk become

\begin{equation}
R\left(h|\epsilon\right)=\frac{R_0\left(h|\epsilon\right)+R_1\left(h|\epsilon\right)}{2}
\end{equation}

We can find optimal threshold $\epsilon_0$ of the classifier $h\left(x\right)$ from $P_{min}=\min_{\epsilon}{R\left(h\middle|\epsilon\right)}$ by combining integral forms with this equation at optimal threshold, we can get

\begin{equation}
P_{min}=\frac{1}{2}\left(\int_{-\infty}^{\epsilon_0}{p_{pos}\left(x\right)}\ dx+\int_{\epsilon_0}^{\infty}{p_{neg}\left(x\right)}\ dx\right)
\end{equation}

which is identical to $P_{min}={OVL}/{2}$ or equivalently, $P_{max}=1-{OVL}/{2}$. The GSSMD considers directional effect of this measure to adapt biological experimental condition. Therefore, the analytical relationship between GSSMD and the maximum probability of success can be defined by 

\begin{equation}
GSSMD=sign\left({\bar{x}}_{pos}-{\bar{x}}_{neg}\right)\times\left(2P_{max}-1\right)
\end{equation}

If the prior probability $P\left(Y=0\right)\neq P\left(Y=1\right)$, GSSMD still correlated with $P_{max}$ but the optimal threshold $\epsilon_0$ may differ from $P\left(Y=0\right)=P\left(Y=1\right)$ case. We may apply Neyman Pearson paradigm to get the optimal threshold $\epsilon_0$ for a given false discovery rate $\alpha$ as recently proposed by Xin Tong \textit{et. al} \cite{tong2018neyman}.

\section{Results}
In this study, Z'-factor\cite{zhang1999simple}, SSMD, robust Z’-factor, robust SSMD, and GSSMD are compared using simulation and RNAi experimental data set to illustrate utility of proposed measure. For all experiments, to estimate the histogram, we used $1+\log2\left(N\right)$ number of bins\cite{cellucci2005statistical}, where $N$ is the number of samples.

\subsection{Normal distribution with equal variance}
\begin{figure}
\centerline{\includegraphics[width = 8cm]{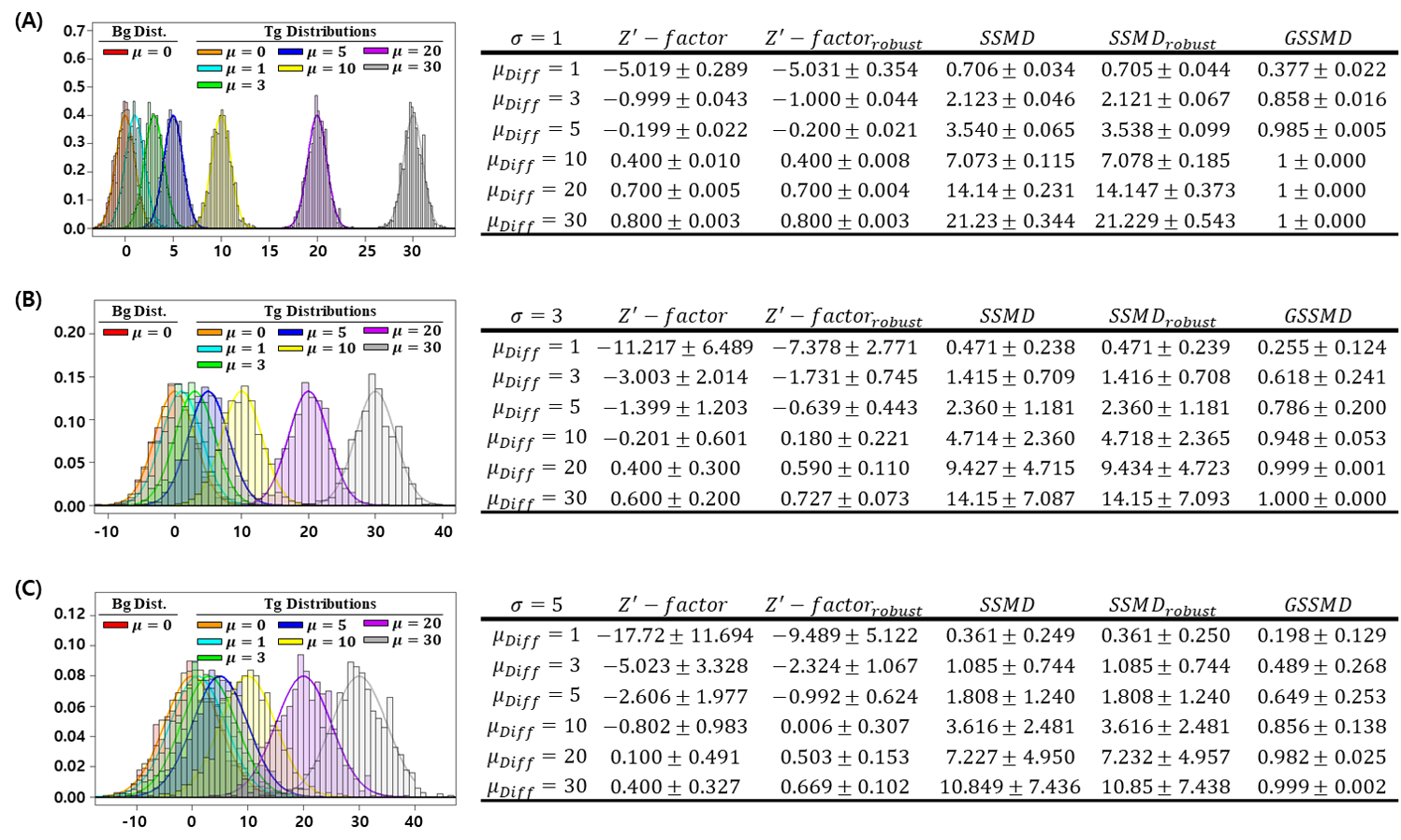}}
\caption{GSSMD provided interpretable and reliable statistical difference estimate in the ideal simulation setup ($\sigma_c=1,\ 3,\ 5\ ,\ n\ =\ 1,000$ in (A), (B), and (C) respectively). The results in each table obtained from $n={10}^6$ independent trials. We sampled $N=1,000$ samples from background distribution and add designated mean difference to get the target distributions. Bg Dist. = Background Distribution, Tg Distributions = Target Distributions.}
\label{fig1}
\end{figure}
In this experiment, three normal distribution cases are compared. In the Fig.~\ref{fig1}, GSSMD increases as the mean difference increases and it decreases as the variance of the distribution increases. Note that in Fig.~\ref{fig1}(A), GSSMD is reaching to it’s maximum (i.e., equals to one) with mean difference of 10 (yellow curve), so there is no significant overlap between the two distributions. In other cases (see Fig.~\ref{fig1}(B) and (C)) due to an increase in distribution variance, the value gradually decreases, indicating an increase in overlap. For detailed comparison with visual illustration and log-normal distribution case study are provided in supplementary material (see Figure S2 and S3). 

\subsection{Effect of outliers}
\begin{figure}
\centerline{\includegraphics[width = 8cm]{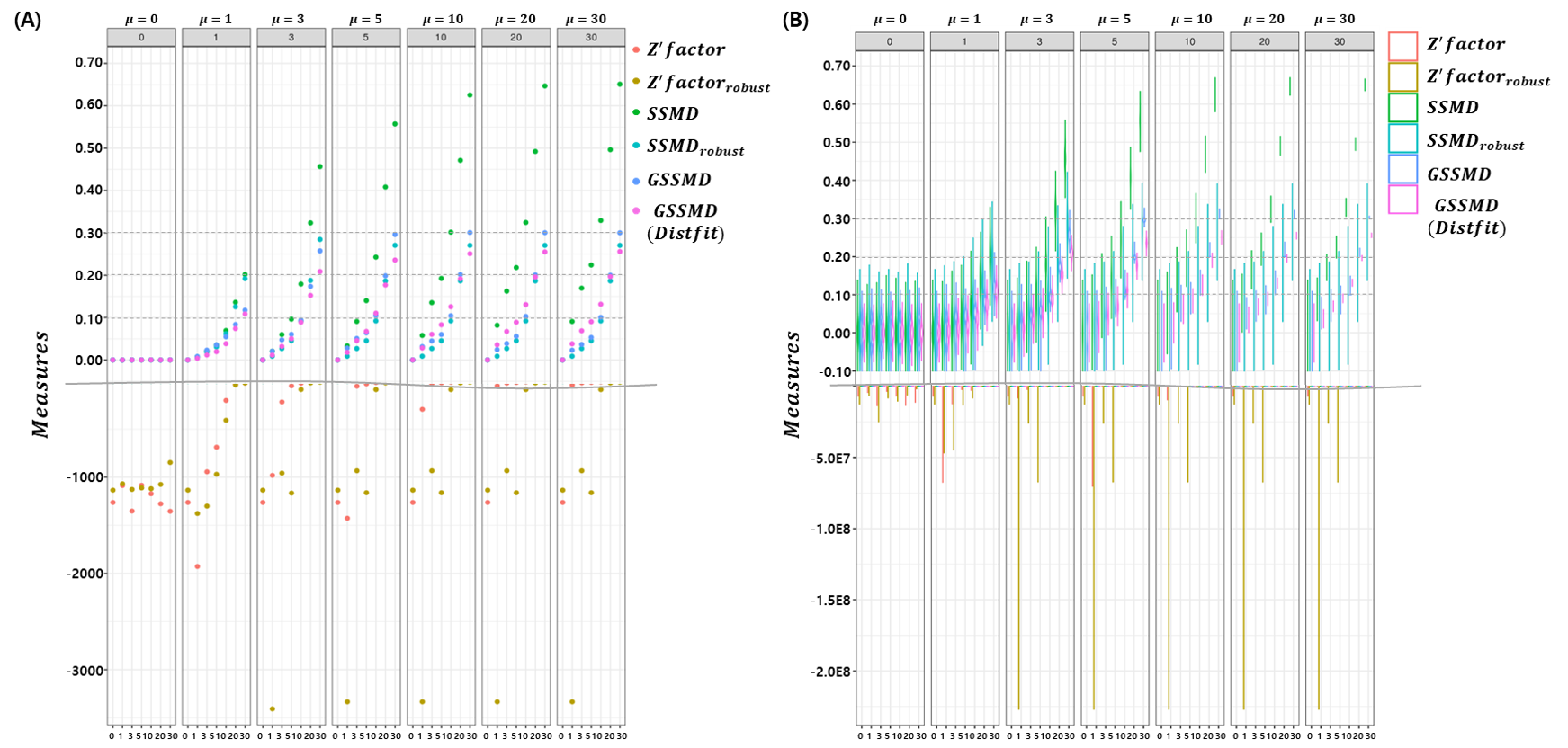}}
\caption{(A) GSSMD is relatively sensitive to outliers compared to Z'-factor and SSMD. Red, green and blue dots represent Z'-factor, SSMD and GSSMD respectively. Robust version of SSMD ${\rm SSMD}_{robust}$ also showed similar performance as GSSMD. The dot represents mean of $n={10}^6$ independent random samplings. (B) Violin plot of (A). GSSMD had the smallest variance in comparison with all other measures and its value converged to the percentage of outlier when $\mu_{diff}\geq5$. The horizontal grey line is inserted to notice scale diffence.}
\label{fig2}
\end{figure}
Unlike ideal simulation case, real experimental measurements are prone to noise and outliers. In this experiment the outlier effect is simulated by replacing positive samples with outlier samples drawn from six different normal distributions. The outliers are generated using $\mathcal{N}\left(\mu,\sigma\right)$, where the variance $\sigma\ =\ 1$ for all distributions. For each case, $N=1,000$ positive and negative samples were generated and $0 - 30\ \%$ positive samples were replaced with outlier samples. We performed $n={10}^6$ independent random trials to estimate mean and variance of each measure. 

Fig.~\ref{fig2} shows the effect of outliers on six measures. In the subplot of Fig.~\ref{fig2}(A), we can see that when the proportion and mean values of the outliers are increased, the mean values of the Z'-factor in various outlier percentile reaches to $-1,000$ even when there is no difference in the population parameters. GSSMD, on the other hand, converges to the percentage of outliers. Especially when the mean of outlier samples is greater than\ 5. In comparison to Z'-factor which is adversely sensitive, SSMD and robust version of SSMD or ${\rm SSMD}_{robust}$ measures are less sensitive to outliers and approaches to specific values when the mean of outlier samples is greater than 10. However, the values of SSMD has no intuitive meaning when distributions of the data are unknown and in additional to this the value of ${\rm SSMD}_{robust}$ varied greatly depending on different sampling scheme (Fig.~\ref{fig2}(B)). Thus, GSSMD is the reliable measure for detecting outliers at small mean differences because its values have intuitive meaning, indicating the percentage of outliers. For the robustness analysis of the proposed method againt different noise conditions see Figure S4 and S5 in supplementary material.

\subsection{Finite sample property of GSSMD}
\begin{figure}
\centerline{\includegraphics[width = 8cm]{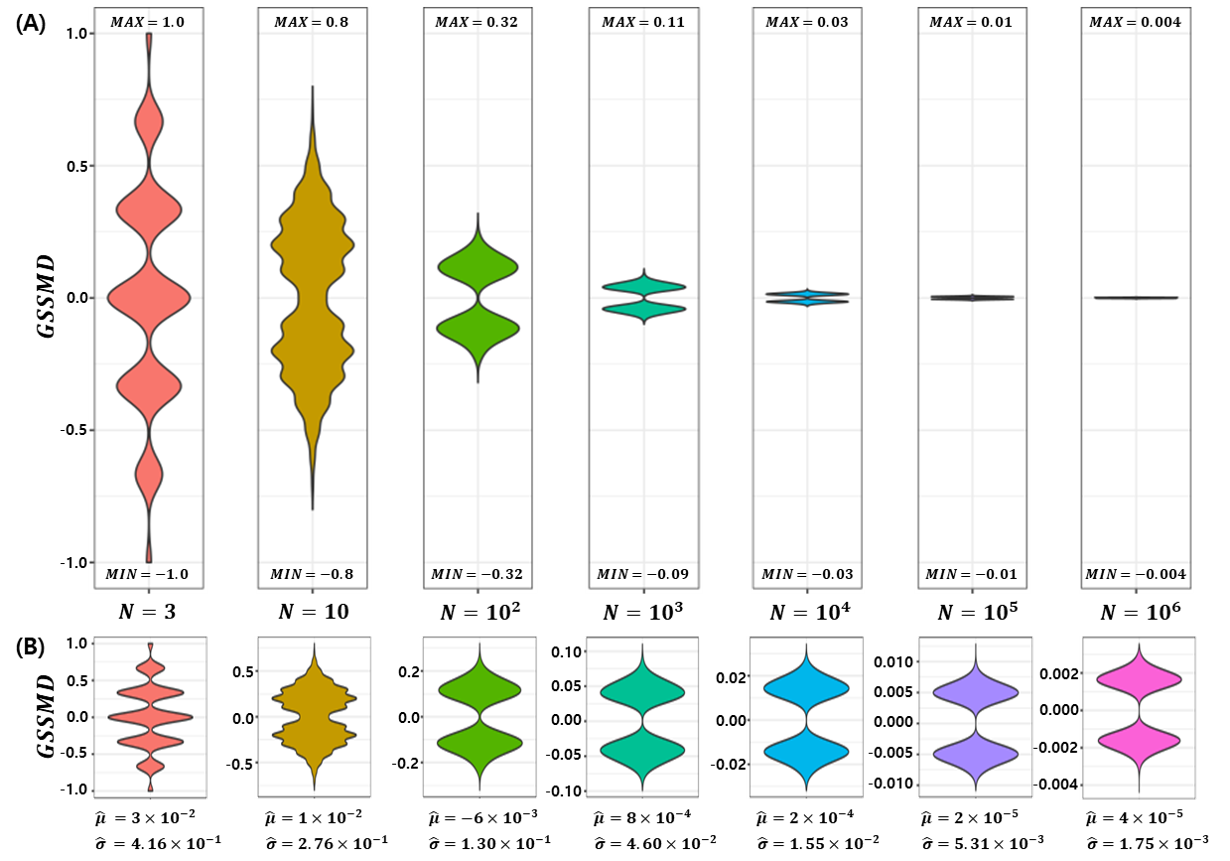}}
\caption{GSSMD score for overlapping distributions, effect of sample-size and lower bound for reliable screening. (A) Violin plot of GSSMD range -1 to 1, (B) Violin plot of GSSMD with matched scale.}
\label{fig5}
\end{figure}
In this experiment, we estimated the lower-bound of GSSMD for two sample sets. In particular, we tested normal and log-normal distribution cases both with $\mu=0$ and $\sigma=1$. Theoretically, the expected GSSMD value for both scenarios is 0, however due to sampling variations the calculated value might be different due to the variations in estimation of PDFs. Therefore, we performed an $10,000$ independent trials and calculated the mean, variance and extreme values of GSSMD for each sample size.

Fig.~\ref{fig5} show the estimated GSSMD for various sample size ($N=3 - {10}^6$). As expected, as the sample size increase, the distance between the mean of the estimated GSSMD and 0 decrease. We also found that the shape of distributions of estimated GSSMD with the sample size of more than 100 is stabilized. The lower-bound of GSSMD for a specific sample size can be seen in the figure. Here we showed only normal distribution case, but the overall behavior is quite similar for log-normal distribution as well. (see supplementary material: Figure S6 and S7). 

In fact, for log-normal distribution the lower-bound is even smaller than the normal distribution case, however for simplicity one can choose the maximum of the normal distribution to avoid possible false positives. This robust and intuitive nature of GSSMD measure make it a suitable choice for reliable assessment of assay quality. To reject null hypothesis, throughout our simulation study, we chose $5\%$ as the threshold of GSSMD for sample size $N=1,000$.

\subsection{Case study}

\begin{figure}
\centerline{\includegraphics[width = 8cm]{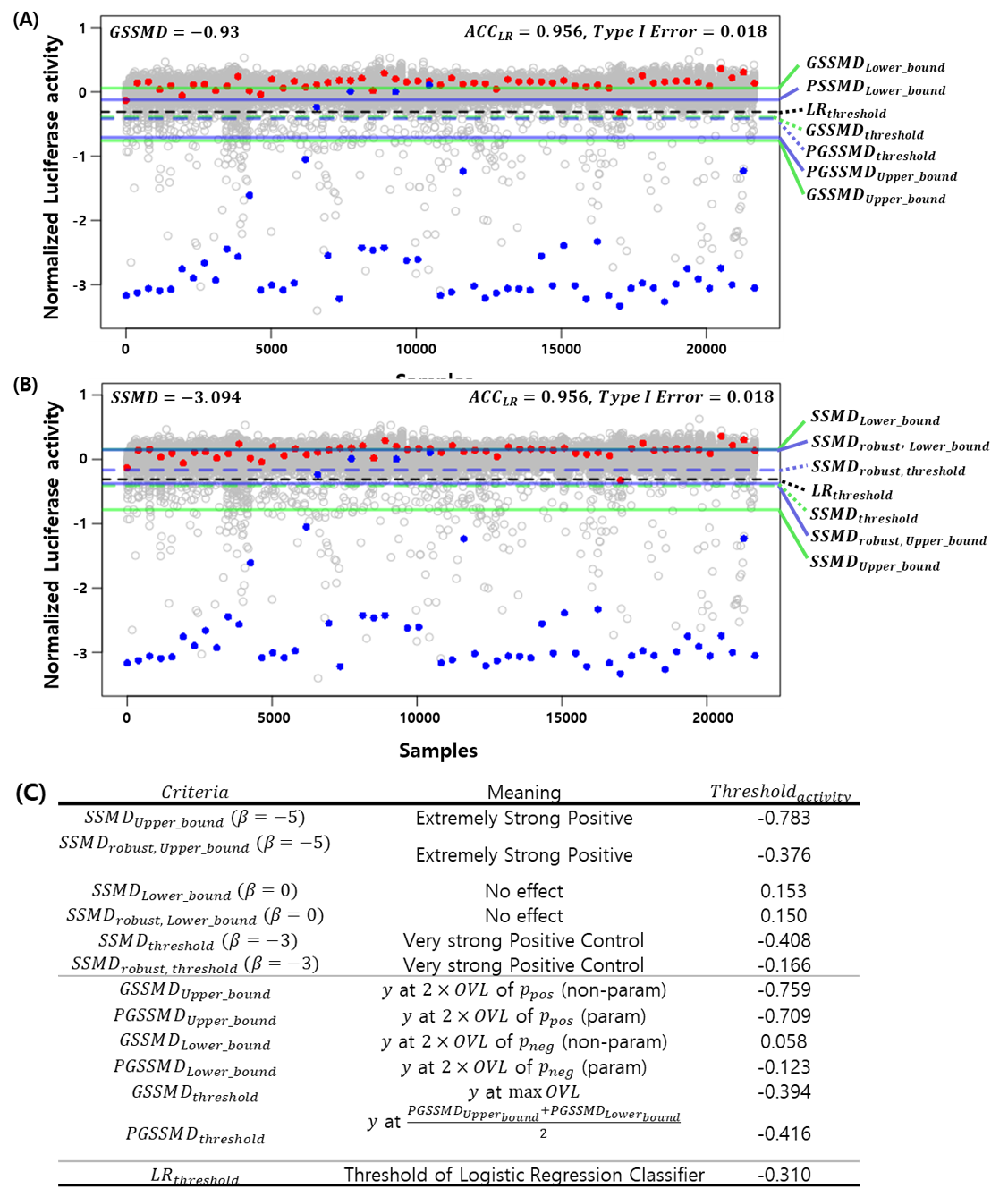}}
\caption{GSSMD can be used to find hit thresholds in RNAi screening. (A) Upper and lower bounds of GSSMD  as well as threshold values. (B) Upper and lower bounds of SSMD and robust SSMD as well as threshold values. (C) Corresponding threshold values for each case. We set the GSSMD threshold at the point where the value of overlap is maximum. Both recommended SSMD and GSSMD threshold were similar to the threshold of optimal logistic regression classifier (Accuracy = $0.956$, Type I error = $0.018$)}
\vspace{-0.15in}
\label{fig6}
\end{figure}

For the case study, we selected RNAi screening of cell viability in Drosophila Kc167 cells\cite{boutros2006analysis} and performed the plate aggregated type of hit selection. Fig.~\ref{fig6} shows the calculated measures on the dataset, where red dots represent negative controls (GFP, Rho, no RNAi treatment) and blue dots represent positive controls (RNAi for D-IAP1 which inducing time-dependent cell death in Drosophila Kc167 cells). Grey dots are samples treated with RNAi, and it is unknown how much reduction in cell viability indicates targeted RNAi binding. As discussed earlier, like SSMD, the proposed GSSMD can also be used to identify thresholds for hit selection. Since the risk of misclassification $R_0\left(h\right)$ dependent on the overlap between two distributions, we can find the optimal threshold of GSSMD for a given FDR based on the value of overlap in a certain range. 

Fig.~\ref{fig6}(A) shows upper and lower bounds of GSSMD as well as threshold values. The upper and lower bounds of each measure are calculated as discussed earlier. In short, we read the value of luciferase activity $y$ at the $OVL$ on each ECDF (positive and negative control distributions, see supplementary material: Figure S12).The threshold was selected by $x$ at maximum $OVL$ occur. In order to evaluate the performance of each measure for finding a correct threshold of hit selection, we trained a logistic regression classifier based on positive and negative control samples. The logistic regression class separation boundary is used as a reference for an optimal threshold of luciferase activity. In particular, we trained a logistic regression classifier using the data from the first plate of the replicated experiment, and tested it on the data of second plate. The threshold provided by the optimal logistic regression classifier (accuracy = $0.956$, type I error = $0.018$) was similar to the GSSMD threshold. ($7$ \% overlap). 

Fig.~\ref{fig6}(B) show thresholds values as well as the upper and lower bound of SSMD and robust SSMD. Since the recommended values of SSMD and robust SSMD are dependent on the type of effect size or strength of positive control, therefore most strong and weak criteria are used as upper and lower bounds. The recommended criteria of SSMD for RNAi screening is 3 and the threshold is close to the threshold of logistic classifier. Threshold for robust SSMD is found to be too loose in this case; however, the lower bound is closest to the optimal classification boundary. Fig.~\ref{fig6}(C) shows corresponding values of luciferase activity for each criterion.

\section{Discussion and Conclusion}
Effect size measures are used in many domains including bioassay as we described above. We showed that non-parametric estimation of effect size based on overlap statistics may provide robust and interpretable estimate of effect size compared to conventional measures. As described in the case study, SSMD may be too loose for suitable selection thresholds for hits even in experiments for which the metric is designed. Moreover, even if there is a guideline to choose alternative threshold, the categorization of positive control is subjective due to the qualitative nature of the guideline. In contrast, the proposed GSSMD measure is relatively sensitive to the detect changes in biological assays. Non-parametric characteristics may cause deviations in the estimates of the measurements, but we have shown that the appropriate number of samples and bin size can minimize this variability.

It may be infeasible to use histogram based overlap estimation in extremely small sample size. In those cases, we can use parametric distribution fitting method based on the understanding of experiment (see supplementary material: Figure S8, S9, and S11). It is also possible to use KDE based estimation instead of histogram based method. We found that when we have enough sample number ($n\geq100$), histogram with adaptive bin-size showed better estimates compared to those KDE based method (see supplementary material: Figure S10). 

 We believe that the proposed method can be easily implemented in many biological applications such as bioassay, gene expression meta analysis etc. The supplemental material and implementation code are available at the author’s Github page (https://github.com/psychemistz/gssmd).

\section*{Acknowledgment}
This research was funded by Center of Excellence in Intelligent Engineering Systems (CEIES), King Abdulaziz University, Jeddah, Saudi Arabia. We thank Professor Alfonso Rodriguez-Molares, Norwegian University of Science and Technology for sharing a MATLAB implementation of GCNR, a similar task performed in ultrasound image quality assessment. We also thank Yoonhyeok Lee for helpful discussion about applying GSSMD measure on the small-scale scenario.

\bibliography{gssmdB383}

\end{document}